\documentclass{article}

\usepackage[nonatbib,preprint]{neurips_2025}

\usepackage{placeins}
\usepackage[utf8]{inputenc} 
\usepackage[T1]{fontenc}    
\usepackage{hyperref}       
\usepackage{url}            
\usepackage{booktabs}       
\usepackage{amsfonts}       
\usepackage{amsmath}
\usepackage{nicefrac}       
\usepackage{microtype}      
\usepackage{xcolor}         
\usepackage[sorting=none,style=numeric-comp]{biblatex}
\usepackage{csquotes}
\usepackage{graphicx}
\usepackage{xspace}
\usepackage{comment}
\usepackage[capitalise]{cleveref}
\hypersetup{
    colorlinks,
    pdfsubject={},
    citecolor=blue,
    urlcolor=blue,
    linkcolor=blue,
}

\newcommand{\ours}{\textsc{Glow}\xspace}

\title{\textsc{Glow}: A Unified Particle Flow Transformer}
\workshoptitle{Machine Learning and the Physical Sciences} 

\author{%
  Dmitrii Kobylianskii\textsuperscript{2,*}, \quad
  Samuel Van Stroud\textsuperscript{1,*}, \quad
  Kwok Yiu Wong\textsuperscript{1}, \quad
  Max Hart\textsuperscript{1}, \\
  \textbf{Etienne Dreyer}\textsuperscript{2}, \quad
  \textbf{Eilam Gross}\textsuperscript{2}, \quad
  \textbf{Gabriel Facini}\textsuperscript{1}, \quad
  \textbf{Tim Scanlon}\textsuperscript{1}
  \AND
  \textsuperscript{1}\normalfont{Centre for Data Intensive Science and Industry, University College London} \\
  \textsuperscript{2}Faculty of Physics, Weizmann Institute of Science \\
  \textsuperscript{*}Authors contributed equally. \\
}

\begin{document}
\maketitle

\begin{abstract}
We present \ours{}, a transformer-based particle flow model that combines incidence matrix supervision from HGPflow with a MaskFormer architecture.
Evaluated on CLIC detector simulations, \ours{} achieves state-of-the-art performance and, together with prior work, demonstrates that a single unified transformer architecture can effectively address diverse reconstruction tasks in particle physics.
\end{abstract}

\section{Introduction}

Modern particle physics detectors produce sparse, multi-modal data that must be reconstructed into sets of particles. Particle flow (PFlow) algorithms address this challenge by combining reconstructed charged particle tracks from tracking detectors with calorimeter energy clusters to determine each particle's type, energy, and direction. Traditional algorithms like Pandora~\cite{Marshall:2015rfa,Thomson:2009rp} rely on expert-designed clustering, matching, and subtraction rules to avoid double counting and maximise resolution; these methods tend to be detector-specific and require extensive manual tuning for new geometries or operating conditions.

We frame particle flow reconstruction as an object-detection task and introduce \ours{}, a transformer-based set-to-set PFlow model that achieves state-of-the-art reconstruction performance. The results demonstrate the effectiveness of unified transformer architectures for particle physics reconstruction.

\section{Related Work}

HGPflow~\cite{dibelloReconstructingParticlesJets2023,Kakati:2024bjf,kakati2025hgpflowextendinghypergraphparticle} treats PFlow as a hypergraph prediction problem with supervised \emph{incidence matrices} that associate detector objects (tracks or clusters) to output particles. For clusters, the matrix defines energy-conserving fractional assignments, from which \emph{proxy} particle kinematics are calculated as weighted sums of associated cluster properties before refinement by a regression network.

MLPF~\cite{mokhtarFinetuningMachinelearnedParticleflow2025,pataImprovedParticleflowEvent2024,pataMachineLearningParticle2023,pataMLPFEfficientMachinelearned2021} frames the problem as a direct, supervised mapping from inputs to particles. Rather than explicit set-to-set reconstruction, MLPF enforces one-to-one associations between input elements and particles, suppressing duplicates when track and cluster originate from the same charged particle at the cost of resolution performance.

MaskFormers~\cite{maskformer,mask2former}, originally developed for image segmentation, have recently been adapted to HEP tasks such as vertexing and tracking~\cite{vanstroudSecondaryVertexReconstruction2024a,vanstroud2024transformerschargedparticletrack}, where local attention provides a novel strategy for scaling to high-occupancy detector data. MaskFormers process unordered input sets to simultaneously identify objects, assign constituents, and estimate object-level properties, making them well-suited for PFlow reconstruction, though this application has not been previously explored.

Object condensation has been proposed for combining calorimeter clustering and regression in a single model~\cite{Kieseler:2020oc,Qasim:2022oc}. It lacks distinct input/output representations and enforces a one-to-one assignment between each input and a single condensation point, limiting shared contributions to multiple reconstructed particles. In a PFlow task, it was outperformed by HGPflow~\cite{dibelloReconstructingParticlesJets2023}.

\section{Model}

\begin{figure*}[!ht]
  \centering
  \includegraphics[width=0.8\textwidth]{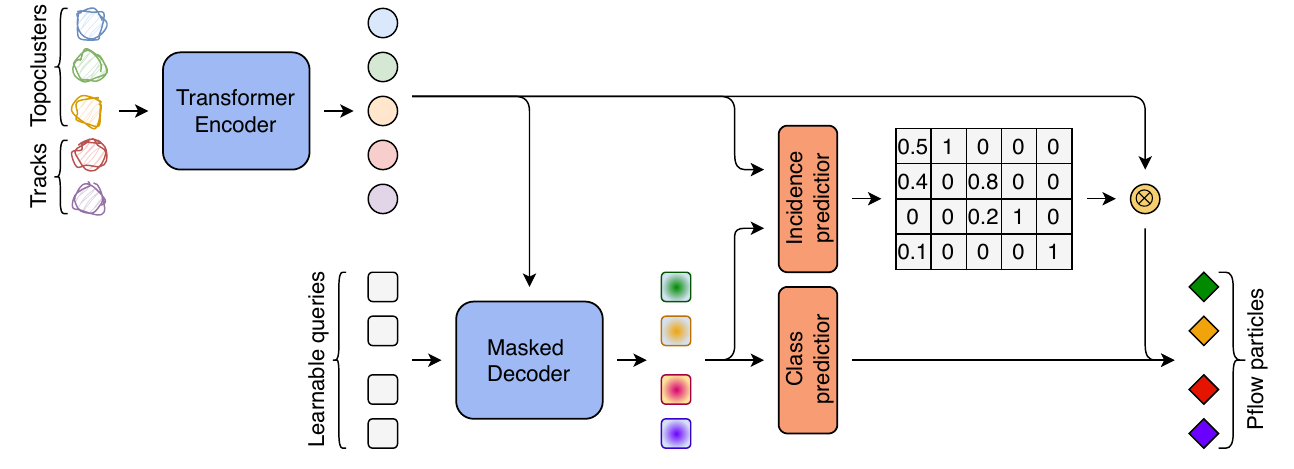}
  \caption{\ours{} architecture overview. Dedicated mask prediction and regression modules are not depicted.}
  \label{fig:diagram}
\end{figure*}

We introduce \ours{}: a unified model combining incidence-based supervision from HGPflow, which enforces energy-consistent associations between detector objects and particles, with MaskFormer's query-based architecture, which naturally handles heterogeneous inputs and enables efficient, permutation-invariant decoding of variable-sized particle sets for PFlow reconstruction (Figure~\ref{fig:diagram}). 
After a six-layer self-attention encoder, learned queries predict unordered particle sets with flexible output cardinality via a four-layer masked cross-attention decoder \cite{mask2former}. Attention masks are dynamically adjusted based on learned similarities between particle queries and input detector objects.
Following HGPflow, the model predicts an incidence matrix $I_{ia}$ representing the fraction of object $i$'s energy attributed to particle $a$, i.e. $I_{ia} = E_{ia} / E_i$, where $E_i$ is the object's total energy. This ensures energy conservation by construction.

The input consists of fully reconstructed tracks and calorimeter topoclusters~\cite{ATLASTopo}, each represented as a feature vector. A fixed number of learnable query embeddings represent particles. At each decoder layer, cross-attention operates between queries and input objects, with attention masks derived from learned similarity scores. 

Each particle query simultaneously predicts: (i) a soft assignment over inputs (approximating the ground-truth incidence matrix), (ii) particle-type labels (photon, electron, hadron), and (iii) kinematic quantities (energy, $p_T$, $\eta$, $\phi$), computed first as incidence-weighted sums and then refined by a regression head.
The training loss combines mask prediction, incidence matrix supervision, classification, and regression objectives. The Hungarian algorithm~\cite{kuhn1955hungarian,10.1145/3442348} matches model predictions to ground truth targets for permutation-invariant training.

This approach enables calibrated particle flow object reconstruction from charged tracks and calorimeter clusters, ensuring differentiability and scalability to future high-granularity detectors.

\section{Experimental Setup}

We use a dataset~\cite{pata_zenodo} of one million simulated $e^+e^- \to$ dijet events at $\sqrt{s} = 380~\mathrm{GeV}$
from the Compact Linear Collider (CLIC) detector~\cite{CLICdp:2017vju, CLICdp:2018vnx}.
This dataset reflects the expected high-multiplicity, high-resolution conditions of future lepton colliders.
The model input consists of fully reconstructed tracks and calorimeter clusters, produced by PandoraPFA~\cite{Marshall:2015rfa}, which applies traditional track finding, clustering, and iterative energy subtraction.
Following~\cite{kakati2025hgpflowextendinghypergraphparticle}, only particles that interact with the detector are included as targets—those with at least one associated track or nonzero calorimeter deposit within $|\eta| < 4$. Converted daughter particles are retained, but targets are defined at the parent level. Each particle is associated with detector objects through the energy-based incidence matrix detailed in~\cite{kakati2025hgpflowextendinghypergraphparticle}.

\ours{} has 12M parameters and was trained for 200 epochs over 23 hours on Isambard-AI~\cite{isambard} using two H100 GPUs. The total batch size was 1024.
Inference is optimized for numerical precision rather than throughput, and reaches approximately 5k events/s per H100.
Code is available on GitHub~\cite{hepattn}.


\FloatBarrier
\section{Results}

For a fair architectural comparison, we retrained a publicly available MLPF model (originally trained on 21.6M events across multiple processes \cite{mokhtarFinetuningMachinelearnedParticleflow2025}) on our dataset, using the original preprocessing from Ref.~\cite{mlpf_code}. The model architecture and hyperparameters were left unchanged. HGPflow results are taken from Ref.~\cite{kakati2025hgpflowextendinghypergraphparticle}.
We evaluate performance on 20k independent dijet events using event- and jet-level metrics. Predictions for all models are processed using the same evaluation pipeline.

\subsection{Event-level performance}

\begin{figure}[htbp]
    \centering
    \includegraphics[width=0.95\linewidth]{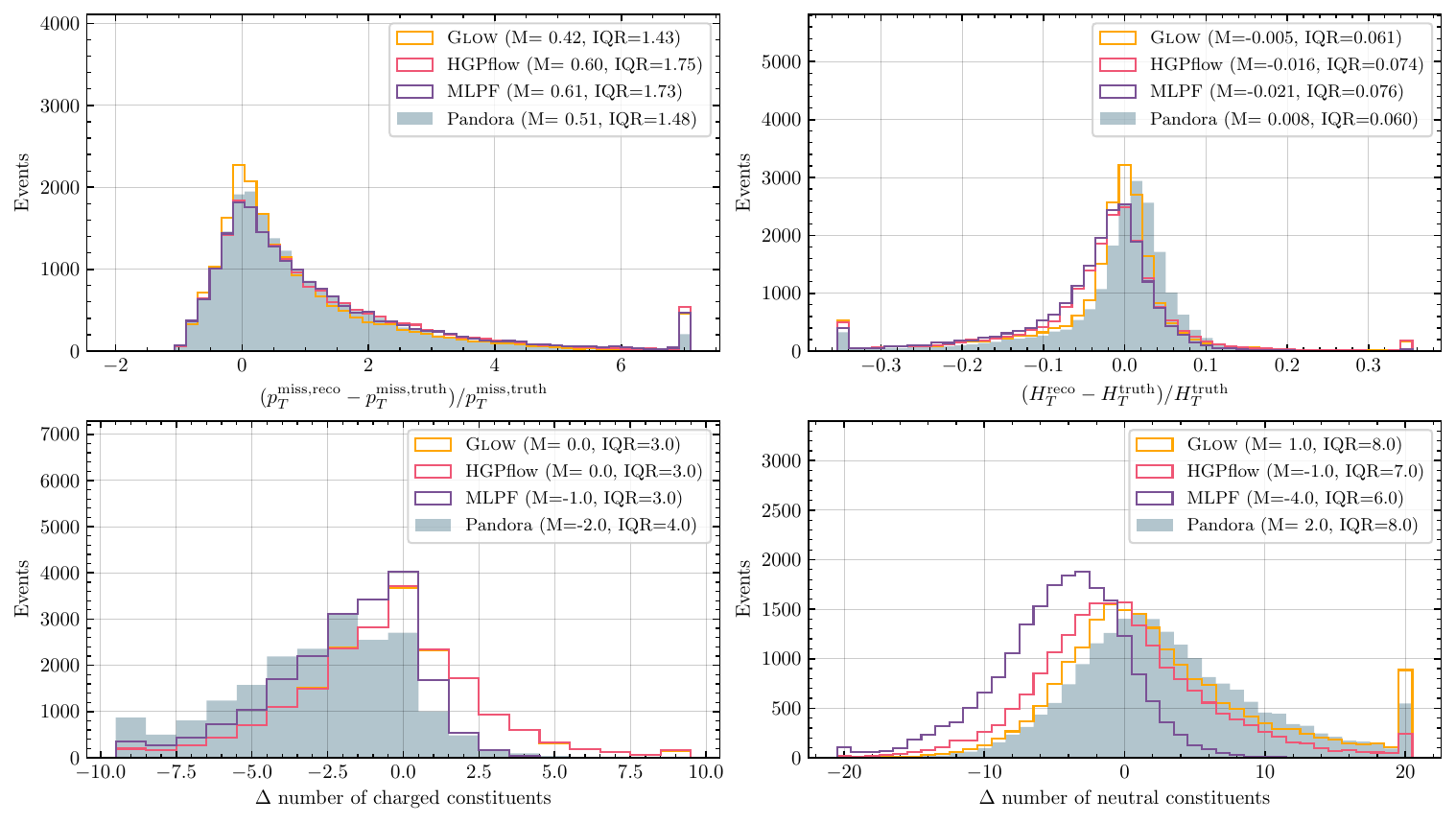}
    \caption{Residual distributions for missing transverse momentum $p_T^{\text{miss}}$ (top left), scalar momentum sum $H_T$ (top right), and counts for charged (bottom left) and neutral (bottom right) particles. The legend includes median and interquartile ranges for each model.}
    \label{fig:event-result}
\end{figure}

\Cref{fig:event-result} shows event-level reconstruction performance across four key metrics. The top panels display residual distributions for missing transverse momentum $p_T^{\text{miss}} = \left|\sum_i \vec{p}_{T,i}\right|$ (left) and scalar transverse momentum sum $H_T = \sum_i \left|\vec{p}_{T,i}\right|$ (right), while the bottom panels show differences between predicted and true particle counts for charged (left) and neutral (right) particles. \ours{} produces the most centered and narrow distributions for both momentum quantities, indicating superior reconstruction accuracy. For particle counting, \ours{} and HGPflow perform similarly due to their shared target definitions. MLPF tends to underestimate the number of constituents. This behavior likely arises from MLPF's target definition, which assigns each topocluster to a single leading particle and can therefore ignore subleading neutrals.

\subsection{Jet-level performance}

\begin{figure}[htbp]
    \centering
    \includegraphics[width=0.95\linewidth]{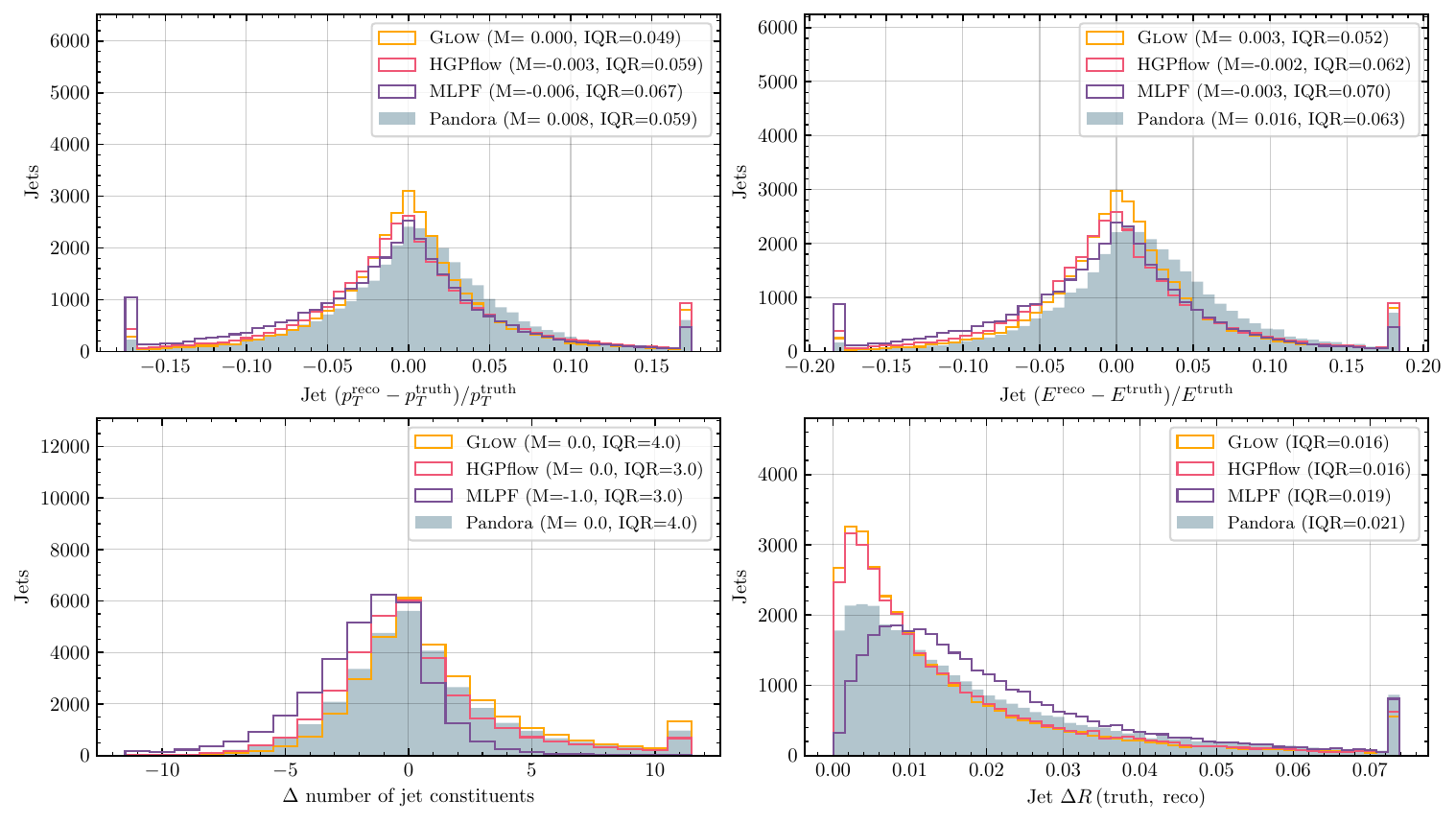}
    \caption{Relative residuals for jet transverse momentum $p_T$ (top left), energy $E$ (top right), number of jet constituents (bottom left), and angular separation $\Delta R$ between reconstructed and truth jets (bottom right).
    The legend includes median and interquartile ranges for each model.}
    \label{fig:jet-residual}
\end{figure}

\begin{figure}[htbp]
    \centering
    \includegraphics[width=0.95\linewidth]{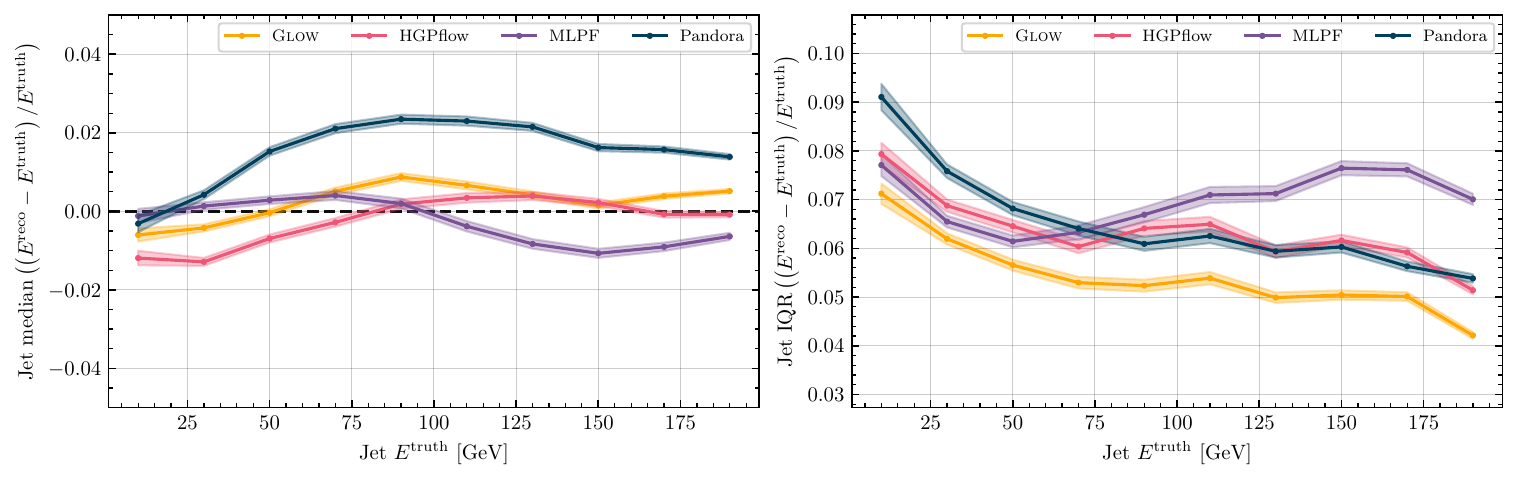}
  \caption{Median (left) and interquartile range (right) of jet energy relative residual distributions as a function of truth jet energy\protect\footnotemark.}
    \label{fig:jet-response}
\end{figure}

We reconstruct jets using FastJet with the generalized $k_T$ algorithm ($R=0.7$), keeping up to two leading jets with $p_T > 10$ GeV. Reconstructed jets are matched to truth jets within $\Delta R < 0.1$.
\cref{fig:jet-residual} shows jet energy and momentum residuals along with angular separation. Both \ours{} and HGPflow outperform MLPF and Pandora. \ours{} produces the narrowest residual distributions and the smallest angular separations to truth jets.
\cref{fig:jet-response} demonstrates how jet energy resolution varies with truth jet energy. \ours{} maintains the best median accuracy (within $\pm 2\%$) and consistently achieves a relative improvement of 15\% in jet energy resolution over HGPflow across all energy ranges. MLPF shows comparable performance below $50$ GeV but degrades at higher energies, as expected from the higher rate of overlapping particles.

\addtocounter{footnote}{-1}

\FloatBarrier
\section{Conclusion}

\footnotetext{The $\pm 1\sigma$ error bands are computed under the assumption of normal distributions using $\sigma_{\text{Median}} = 0.93 \cdot \text{IQR}/\sqrt{N}$ and $\sigma_{\text{IQR}} = 1.16 \cdot \text{IQR}/\sqrt{N}$.}

We present \ours{}, a transformer-based particle flow reconstruction model that combines a MaskFormer architecture with incidence matrix supervision from HGPflow. \ours{} advances particle flow reconstruction  beyond HGPflow by replacing iterative hypergraph refinement with a masked decoder in a fully transformer-based architecture. By avoiding backpropagation through time, training and inference are accelerated, while multi-head cross-attention provides greater expressiveness than incidence-weighted message passing. The use of standard transformer blocks also eases deployment with ONNX or TensorRT. Compared with MLPF, the physics-based incidence matrix formulation allows topoclusters to contribute to multiple particles while preserving energy consistency through fractional assignments.

Evaluated on CLIC detector simulations, \ours{} achieves state-of-the-art reconstruction performance across event- and jet-level metrics, improving jet energy resolution by roughly 15\% relative to HGPflow while reducing biases to within $\pm 2\%$. Together with prior work, we demonstrate that a unified encoder-decoder transformer can effectively handle core reconstruction tasks in particle physics, providing a flexible and scalable foundation for reconstruction at future colliders.
\clearpage

\section*{Acknowledgments}

We gratefully acknowledge the support of the UK's Science and Technology Facilities Council (STFC). S.V.S is supported by ST/X005992/1. M.H., K.Y.W, N.P. are supported by the STFC UCL Centre for Doctoral Training in Data Intensive Science  (ST/W00674X/1) and by departmental and industry contributions. G.F and T.S. receives support from the STFC (ST/W00058X/1) and T.S. is supported by the Royal Society (URF/R/180008).

E.G., E.D., and D.K. are supported by the Minerva Stiftung through funding from the German Federal Ministry of Education and Research under grant number 715027, and the Weizmann Institute for Artificial Intelligence grant program.

We also extend our thanks to UCL for the use of their high-performance computing facilities, with special thanks to Edward Edmondson for his expert management and technical support.

The authors acknowledge the use of resources provided by the Isambard-AI National AI Research Resource (AIRR). Isambard-AI is operated by the University of Bristol and is funded by the UK Government’s Department for Science, Innovation and Technology (DSIT) via UK Research and Innovation; and the Science and Technology Facilities Council [ST/AIRR/I-A-I/1023].
\printbibliography
\end{document}